\documentclass{ws-ijmpe}
\usepackage[super,compress]{cite}
\begin{document}

\markboth{M. R. Hadizadeh, M. Radin, S. Bayegan}{Comment on "Three--dimensional study of the six--body bound--state for the case of effective ..."}

\catchline{}{}{}{}{}
%%%%%%%%%%%%%%%%%%%%%%%%%%%%%%%%%%%%%%%%%%%%%%%%%%%%%%%%%%%%%%%%%%%%

\title{Comment on "Three--dimensional study of the six--body bound--state for the case of effective three-body configuration model" [Int. J. Mod. Phys. E 25, 9 (2016) 1650072].}

\author{M. R. Hadizadeh}%\footnote{Typeset names in
%10~pt Times roman, uppercase. Use the footnote to indicate
%the present or permanent address of the author.}}

\address{Institute for Nuclear and Particle Physics and Department of Physics and Astronomy, 
Ohio University, Athens, OH 45701, USA.}
\address{College of Science and Engineering, Central State University, \\ Wilberforce,
OH 45384, USA.}

%\address{University Department, University Name, Address\\ City, State ZIP/Zone, Country\footnote{State completely without abbreviations, the affiliation and mailing address, including country. Typeset in 8~pt Times italic.}\\ first\_author@university.edu}

\author{M. Radin}
\address{Department of Physics, K. N. Toosi University of Technology, Tehran, Iran}
%second\_author@group.com}

\author{S. Bayegan}
\address{Department of Physics, University of Tehran, Tehran, Iran}

\maketitle

\begin{history}
\received{Day Month Year}
\revised{Day Month Year}
%\accepted{Day Month Year}
%\comby{(xxxxxxxxxx)}
\end{history}

\begin{abstract}
The authors argue that the calculated $^6$He binding energies by the solution of the coupled Faddeev--Yakubovsky integral equation in a Three--dimensional scheme reported by E. Ahmadi Pouya and A. A. Rajabi [Int. J. Mod. Phys. E 25, 9 (2016) 1650072] are incorrect. The formalism of the paper has serious mistakes and the numerical results are quite misleading because such a  calculation even with small grids for Jacobi momenta and the angle variables leads to a huge memory of 37.8 PB (petabyte) which cannot even be achieved on present supercomputers.
\end{abstract}

\keywords{Faddeev--Yakubovsky equations; Six--body bound state; Three--dimensional approach.}

\ccode{PACS numbers: 21.45.-v, 21.10.Dr, 27.20.+n}

\vskip1cm

%\tableofcontents

Ahmadi Pouya and Rajabi have recently studied the halo structure of $^6$He as a six--nucleon bound state. 
As it is shown in Ref. \cite{Witala-FBS51} the formulation of the Faddeev--Yakubovsky equations for a six--body bound state leads to five coupled equations, which can be reduced to two coupled equations for two--neutron halo nuclei, like $^6$He.
In Ref. \cite{Ahmadi-IJMPE25} Ahmadi Pouya and Rajabi have presented the formulation of the six--body Faddeev--Yakubovsky equations in a Three--Dimensional (3D) approach, without using a partial wave decomposition.
The paper has serious problems and mistakes in its structure, not only in the formalism but also in the numerical implementation.
In the following, we have addressed few of these mistakes and flaws.

\section{Mistakes in the formalism}

\begin{enumerate}
\item Each Yakubovsky component $K$ and $H$ depends on 5 Jacobi momentum vectors. By choosing any coordinate system and setting one of the vectors parallel to the $z-$axis and another vector in the $x-z$ plane, the number of independent variables for each Yakubovsky component is 12, including:

\begin{itemize}
\item 5 variables for the magnitude of 5 Jacobi momentum vectors
\item 4 variables for the spherical angles
\item 3 variables for the azimuthal angles
\end{itemize}

So, the representation of Yakubovsky components in Eqs. (4.3) and (4.4) is incorrect and consequently, the integral equations (4.7) and (4.8) are incorrect.

\item How it is possible to have the two--body $t-$matrices outside of the kernel of Faddeev--Yakubovsky integral equation in a 3D scheme?
Clearly the representation of two--body $t-$matrices in the integral equations (4.7) and (4.8) is incorrect and is inconsistent with the three-- and four--body formalism in a 3D representation \cite{Elster_PRC58,Schadow_FBS28,Liu_FBS33,Fachruddin_MPLA18,FachuddiN_{jac}RC68,Liu_PRC72,LiN_{jac}RC76,LiN_{jac}LB660,LiN_{jac}RC78,Gloeckle_EPJA43,3d-0,3d-1,3d-2,3d-3,3d-4,3d-5,3d-6,3d-7,3d-8,3d-9,3d-10}.
The only way to exclude the two--body $t-$matrices from the kernel of Faddeev--Yakubovsky integral equations is using the one--term separable potential in the $s-$wave channel \cite{pw-1, pw-2} which is not applicable here, because the formalism is given for a general interaction in a 3D representation. 

\item Other obvious mistakes in the formalism and mainly in the Eqs. (4.7) and (4.8) are:

\begin{itemize}
\item There is no angular dependence in the $K$ and $H$ Yakubovsky components appeared in the kernel of three--dimensional integral equations, which is inconsistent with the definitions given in Eqs. (4.3) and (4.4).

\item How it is possible that $K$ component on the left--hand side of the integral equation (4.7) be dependent to $x_{32'}$ and $\phi_{42'}$, whereas the $x_{32'}$ and $\phi_{42'}$ are the integration variables?

\item How it is possible that $H$ component on the left--hand side of the integral equation (4.8) be dependent to $X_{32'}$ and $\Phi_{42'}$, whereas the $X_{32'}$ and $\Phi_{42'}$ are the integration variables?

\item How it is possible to have the $4\pi$ factor in Eqs. (4.7) and (4.8) in a 3D representation? This factor just appears in a partial wave representation.

\item How the two--body $t-$matrices $t(a_1,\pi_2,x;\epsilon)$ and $t(b_1,b'_2,y;\epsilon^*)$ are dependent on the angle variables $x$ and $y$, whereas there is no angular dependence to the angles $x$ and $y$ in the $K$ and $H$ Yakubovsky components?

\item How the two--body $t-$matrix $t(a_1,\pi_2,x;\epsilon)$ is outside of the kernel of integral equation (4.7), whereas the shifted momentum $\pi_2$ is dependent on the integration variable $a'_2$?

\item How the two--body $t-$matrix $t(b_1,b'_2,y;\epsilon^*)$ is outside of the kernel of integral equation, whereas it is dependent on the integration variable $b'_2$?

\end{itemize}

\end{enumerate}

In summary, the published formalism has mistakes which can completely change the results of the calculation. The mistakes in the formalism can be easily verified by simplification of the problem to a four-- or three--body bound state.
Clearly, the published formalism cannot reproduce the 3D representation of Faddeev and Yakubovsky equations given in Refs. \cite{3d-0} and \cite{3d-1}.

\section{Numerical challenges}
\begin{enumerate}

\item Let us have a look at the dimension of the problem, discussed in section 5 of the paper. As we have discussed in item 1 of section 1 of this comment, each Yakubovsky component depends on 12 variables, not 10 variables.
So, the size of the problem given in Eq. (5.1) of the paper is incorrect.
The size of the coupled Yakubovksy components would be: 
\begin{equation}
N_{jac}^5  \cdot N_{sph}^4  \cdot N_{pol}^3  \cdot 2 = 20^5  \cdot 14^4   \cdot 14^3   \cdot 2  =  6.75 \cdot 10^{14},
\end{equation}
where:
\begin{itemize}
\item $N_{jac}$ is the number of mesh points for the magnitude of Jacobi momenta,
\item $N_{sph}$ is the number of mesh points for the spherical angles,
\item $N_{pol}$ is the number of mesh points for the azimuthal angles.
\end{itemize}

If the authors have used at least 7 iterations, the total size of the problem would be $4.72  \cdot 10^{15}$ which is equal to $3.78 \cdot 10^{16}$ byte \footnote{A double-precision variable occupies 8 bytes in computer memory.}, or about $37.8$ PB (petabyte)!
This huge memory is about $28.3$ times bigger than the memory of Sunway TaihuLight supercomputer (with 1.31 PB memory), which is ranked number one in the TOP500 list as the fastest supercomputer in the world \cite{TaihuLight}.
Since such a supercomputer with the huge memory of $37.8$ PB doesn't exist, we wonder how the authors have done this super expensive calculation. %, of course, if they have done any!
If we even consider the wrong size given in Eq. (5.1) of the paper, the required memory just for two Yakubovsky components would be about $192756$ GB (Gigabyte) which is not accessible in many supercomputers, and clearly is much more bigger than the memory of IPM HPC cluster \cite{HPC_IPM} which the authors have used in their calculations.

\item The starting vectors for the iteration procedure given in Eqs. (5.3) and (5.4) are incorrect. They should be dependent on 5 Jacobi momenta, whereas the dependency of Yakubovsky components to the first Jacobi momenta $a_1$ and $b_1$ is not considered.
 
\item The calculated two--body $t-$matrices given in Eqs. (5.5) and (5.6) is just valid for a one--term separable potential, like Yamaguchi potential. 
For a general case, the two--body $t-$matrices in a 3D representation should be obtained by the solution of three--dimensional Lippmann-Schwinger integral equations given in Ref. \cite{Elster_FBS24}.
The authors have referred to Ref. [33] in their paper for the calculation of two--body $t-$matrices in 3D scheme which is quite irrelevant.
So, it is not clear how the authors have calculated 3D form of two--body $t-$matrices for Baker, Volkov and Malfliet--Tjon potentials which are not one--term separable potentials.

\item The Fourier transformation of Gauss--type Baker and Volkov potentials given in Eq. (7.4) and (7.6) are incorrect. The results should be a Gaussian potential.

\item There is no information about the momentum cutoffs in the numerical solution of the coupled Yakubovsky equation. How the authors are convinced that the obtained binding energies are cutoff independent, when they have used only 20 mesh points for the magnitude of Jacobi momenta?

\item There is an obvious mistake in the evaluation of the permutation operator $P$ given in Eq. (A.4) of the paper which leads to another mistake in derivation of Eq. (3.6). We address the authors to Eq. (2.9) of Ref. \cite{3d-0} for the correct evaluation of operator $P$.  

\end{enumerate}

There are many other issues in the derivation of Yakubovsky equations and also in the numerical implementations that we have not discussed here. But we believe the above--mentioned mistakes are quite enough to ensure us that the authors have reported not genuine results.
Not only the formalism of the paper has serious mistakes, but also there is no supercomputer to handle such a super expensive calculation and undoubtedly the authors have not been able to perform such a calculation.
 
%Therefore the paper should be rejected not only based on the scientific and technical quality but also based on ethical grounds.

%\section*{Acknowledgements}


\begin{thebibliography}{0}

\bibitem{Witala-FBS51}
H. Wita{\l}a and W. Gl\"ockle, Few-Body Syst. {\bf 51}, 27 (2011).

\bibitem{Ahmadi-IJMPE25}
E. Ahmadi Pouya and A. A. Rajabi, Int. J. Mod. Phys. E {\bf 25}, 1650072 (2016).

\bibitem{Elster_PRC58}
Ch. Elster, W. Schadow, H. Kamada, W. Gl\"ockle, Phys. Rev. C {\bf 58}, 3109 (1998).

\bibitem{3d-0}
Ch. Elster, W. Schadow, A. Nogga, W. Gl\"ockle, Few Body Syst. {\bf 27}, 83 (1999).

\bibitem{Schadow_FBS28}
W. Schadow, Ch. Elster, W. Gl\"ockle, Few Body Syst. {\bf 28}, 15 (2000).

\bibitem{Liu_FBS33}
H. Liu, Ch. Elster, W. Gl\"ockle, Few Body Syst. {\bf 33}, 241 (2003).

\bibitem{Fachruddin_MPLA18}
I. Fachruddin, Ch. Elster, W. Gl\"ockle, Mod. Phys. Lett. A {\bf 18} 452 (2003).

\bibitem{FachuddiN_{jac}RC68}
I. Fachuddin, Ch. Elster, W. Gl\"ockle, Phys. Rev. C {\bf 68}, 054003 (2003).

\bibitem{Liu_PRC72}
H. Liu, Ch. Elster, W. Gl\"ockle, Phys. Rev. C {\bf 72}, 054003 (2005).

\bibitem{LiN_{jac}RC76}
T. Lin, Ch. Elster, W. N. Polyzou, W. Gl\"ockle, Phys. Rev. C {\bf 76}, 014010 (2007).

\bibitem{3d-1}
M. R. Hadizadeh and S. Bayegan, Few Body Syst. {\bf 40}, 171 (2007) 

\bibitem{LiN_{jac}LB660}
T. Lin, Ch. Elster, W.N. Polyzou, W. Gl\"ockle, Phys. Lett. B {\bf 660}, 345 (2008).

\bibitem{LiN_{jac}RC78}
T. Lin, Ch. Elster, W.N. Polyzou, H. Wita{\l}a, W. Gl\"ockle, Phys. Rev. C {\bf 78}, 024002 (2008).

\bibitem{3d-2}
M. R. Hadizadeh and S. Bayegan, Eur. Phys. J. A {\bf 36}, 201 (2008) 

\bibitem{3d-3}
S. Bayegan, M. R. Hadizadeh and M. Harzchi, Phys. Rev. C {\bf 77}, 064005 (2008) 

\bibitem{3d-4}
S. Bayegan, M. R. Hadizadeh and W. Gl\"oeckle, Prog. Theor. Phys. {\bf 120}, 887 (2008) 

\bibitem{3d-5}
M. R. Hadizadeh and S. Bayegan, Mod. Phys. Lett. A {\bf 24}, 816 (2009) 

\bibitem{Gloeckle_EPJA43}
W. Gl\"ockle, I. Fachruddin, Ch. Elster, J. Golak, R. Skibinski, H. Wita{\l}a, Eur. Phys. J. A {\bf 43}, 339 (2010).

\bibitem{3d-6}
M. R. Hadizadeh, L. Tomio and S. Bayegan, AIP Conf. Proc. {\bf 1265}, 84 (2010)

\bibitem{3d-7}
M. R. Hadizadeh, L. Tomio and S. Bayegan, Phys. Rev. C {\bf 83}, 054004 (2011)

\bibitem{3d-8}
M. R. Hadizadeh and L. Tomio, AIP Conf. Proc. {\bf 1423}, 136 (2012) 

\bibitem{3d-9}
M. R. Hadizadeh, Prog. Theor. Exp. Phys. {\bf 043D01} (2014) 

\bibitem{3d-10}
M. R. Hadizadeh, Ch. Elster, W. N. Polyzou, Phys. Rev. C {\bf 90}, 054002 (2014)


\bibitem{pw-1}
M. R. Hadizadeh, M. T. Yamashita, A. Delfino, L. Tomio and T. Frederico, PoS (XXXIV BWNP) 034 (2012).

\bibitem{pw-2}
M. R. Hadizadeh, M. T. Yamashita, A. Delfino, L. Tomio and T. Frederico, Phys. Rev. A {\bf 85}, 023610 (2012).

\bibitem{TaihuLight}
https://www.top500.org

\bibitem{Elster_FBS24}
Ch. Elster, J.H.Thomas, W. Gl\"ockle, Few Body Syst. {\bf 24}, 55 (1998)

\bibitem{HPC_IPM}
http://math.ipm.ac.ir/hpccluster/

\end{thebibliography}
\end{document}